\documentclass[preprint,prd,aps,showpacs,showkeys,nofootinbib]{revtex4}
\usepackage{graphicx}
\begin{document}


\title{Muon conversion to an electron in nuclei in the $B-L$ symmetric SSM}

\author{Ze-Ning Zhang$^{1}$\footnote{zn\_zhang\_zn@163.com},
Hai-Bin Zhang$^{1,2}$\footnote{Corresponding author.\\hbzhang@hbu.edu.cn},
Xing-Xing Dong$^{1,2}$,
Jin-Lei Yang$^{3,4}$,
Wei Li$^{1}$,
Zhong-Jun Yang$^{5}$,
Tong-Tong Wang$^{1}$,
and Tai-Fu Feng$^{1,2,5,6}$\footnote{fengtf@hbu.edu.cn}}

\affiliation{$^1$Department of Physics, Hebei University, Baoding, 071002, China\\
$^2$Key Laboratory of High-Precision Computation and Application of Quantum Field Theory of Hebei Province, Baoding, 071002, China\\
$^3$CAS Key Laboratory of Theoretical Physics, Institute of Theoretical Physics, Chinese Academy of Sciences, Beijing 100190, China \\
$^4$School of Physical Sciences, University of Chinese Academy of Sciences, Beijing 100049, China \\
$^5$College of Physics, Chongqing University, Chongqing 400044, China\\
$^6$Department of Physics, Guangxi University, Nanning 530004, China}

\begin{abstract}
 In a few years, the COMET experiment at J-PARC and the Mu2e experiment at Fermilab will probe the $\mu-e$ conversion rate in the vicinity of $\mathcal{O}(10^{-17})$ for an Al target with high experimental sensitivity. Within the framework of the minimal supersymmetric extension of the Standard Model with local $B-L$ gauge symmetry (B-LSSM), we analyze the lepton flavor violating (LFV) process of $\mu-e$ conversion in nuclei. Considering the constraint of the experimental upper limit of the LFV rare decay $\mu\rightarrow e\gamma$, the $\mu-e$ conversion rates in nuclei within the B-LSSM can achieve $\mathcal{O}(10^{-12})$, which is 5 orders of magnitude larger than the future experimental sensitivity at the Mu2e and COMET experiments and may be detected in the near future.

\end{abstract}


\maketitle

\section{Introduction\label{sec1}}
In search of new physics (NP) beyond the Standard Model (SM), we previously studied lepton flavor violating (LFV) decays  $l_j\rightarrow l_i\gamma$, $l_j\rightarrow 3l_i$, and  $h\rightarrow l_il_j$ in the minimal supersymmetric extension of the Standard Model with local $B-L$ gauge symmetry (B-LSSM)~\cite{50,Zhang:2021nzv}. In order to further study lepton flavor violating decay processes, here we investigate muon conversion to an electron in nuclei in the B-LSSM. The present  upper limit of the $\mu-e$ conversion rate in Ti nuclei is ${\rm{CR}}(\mu\rightarrow e:\rm Ti)<4.3\times 10^{-12}$ at 90\% confidence level (C.L.)~\cite{CRTi}, and the future experimental sensitivity of ${\rm{CR}}(\mu\rightarrow e:\rm{Ti})$ will be $\mathcal{O}(10^{-18})$~\cite{Barlow:2011zza}.
For $\mu-e$ conversion in nuclei, the best upper limit is ${\rm{CR}}(\mu\rightarrow e:\rm{Au}) < 7\times 10^{-13}$ (90 \% C.L.), which is given by the SINDRUM-II experiment~\cite{SINDRUMII:2006dvw}. The COMET experiment at J-PARC and the Mu2e experiment at Fermilab are next-generation experiments for $\mu-e$ conversion in nuclei. In a few years, both Mu2e at FNAL~\cite{Mu2e:2014fns} and COMET at J-PARC~\cite{COMET:2018auw} are expected to probe the $\mu-e$ conversion rate in the vicinity of $\mathcal{O}(10^{-17})$ for an Al target with high experimental sensitivity~\cite{CGroup:2022tli}.

The LFV decays are forbidden in the Standard Model~\cite{Harnik}. But they can easily occur in new physics models beyond the SM. The $\mu-e$ conversion rate has been calculated in the literature for various extensions of the SM; for instance, seesaw models with right-handed neutrinos~\cite{Riazuddin:1981hz,Chang:1994hz,Ioannisian:1999cw,Pilaftsis:2005rv,Deppisch:2005zm,
Ilakovac:2009jf,Deppisch:2010fr}, scalar triplets~\cite{Raidal:1997hq,Ma:2000xh,Dinh:2012}, fermion singlets~\cite{Sun:2013kga}, and fermion triplets~\cite{Abada:2008ea} can get the $\mu-e$ conversion rate close to the experimental sensitivity. There are some studies for $\mu-e$ conversion in models of supersymmetry (SUSY), such as, the minimum supersymmetric Standard Model (MSSM) \cite{Hisano}, R-parity violating SUSY \cite{Sato}, low-scale seesaw models of minimal supergravity \cite{Ilakovac2013}, the $\mu$ from the $\nu$ supersymmetric Standard Model~\cite{Zhang:2013jva,Zhang2}, the MSSM with local gauged baryon and lepton number~\cite{Guo:2018qhv}, and the minimal R-symmetric supersymmetric standard model~\cite{Sun:2020puo}. For muon conversion to electron conversion in nuclei, there are also some studies in models of non-SUSY, for instance, the unparticle model \cite{Ding,sksup}, the littlest Higgs model \cite{Blanke,Aguila}, left-right symmetric models \cite{Bonilla}, the 331 model \cite{Huong}, and so on. In this work, we analyze the LFV process $\mu-e$ conversion in nuclei within the B-LSSM.

The gauge symmetry group of the B-LSSM~\cite{5,6,46,47,48,49,B-L1,B-L2} extends that of the MSSM~\cite{MSSM,MSSM1,MSSM2,MSSM3,MSSM4} to $SU(3)_C\otimes SU(2)_L\otimes U(1)_Y\otimes U(1)_{B-L}$, where $B$ stands for the baryon number and $L$ for the lepton number. The B-LSSM can provide many more candidates for dark matter compared to the MSSM, for example, new neutralinos corresponding to the gauginos of $U(1)_{B-L}$, additional Higgs singlets, and sneutrinos~\cite{16,1616,DelleRose:2017ukx,DelleRose:2017uas}. In the B-LSSM, magnetic and electric dipole moments of leptons and quarks have been analyzed~\cite{MDM-1,MDM-2,MDM-3}.

The present experimental upper limit on the LFV branching ratio of $\mu\rightarrow e\gamma$ at the MEG experiment is given as~\cite{AMB44}
\begin{eqnarray}
&&{\rm{Br}}(\mu\rightarrow e\gamma)<4.2\times10^{-13}.\label{a4}
\end{eqnarray}
The best upper limit on the LFV decays for the branching ratio of $\mu\rightarrow e\gamma$ can give a large constraint on the parameter space  in the B-LSSM, compared to the other LFV decays $\mu\rightarrow 3e$  and  $h\rightarrow e \mu$~\cite{50,Zhang:2021nzv}.
In this paper, the LFV process $\mu-e$ conversion rates in Ti, Au, and Al targets will be analyzed in the B-LSSM, considering the constraint of the present experimental limits on the branching ratio of $\mu\rightarrow e\gamma$.

The paper is organized as follows. In Sec. II, we mainly introduce the B-LSSM including its superpotential and the general soft breaking terms. In Sec. III, we give an analytic expression for the $\mu-e$ conversion rates in nuclei in the B-LSSM. In Sec. IV, we give the numerical analysis, and the summary is given in Sec. V. Finally, some tedious formulas are collected in the appendixes.

\section{B-LSSM\label{sec2}}
The B-LSSM is one of the extended models of the MSSM. Compared with the MSSM, the B-LSSM~\cite{46,47,48,49,B-L1,B-L2} adds two singlet Higgs fields $\hat{\eta}_{1}\sim(1,1,0,-1)$ and $\hat{\eta}_{2}\sim(1,1,0,1)$ and three generations of right-handed neutrinos $\hat \nu^c_i \sim(1,1,0,1/2)$. The gauge symmetry group of the B-LSSM is $SU(3)_C\otimes SU(2)_L\otimes U(1)_Y\otimes U(1)_{B-L}$. At the same time, the other chiral superfields and their quantum numbers are given as
\begin{eqnarray}
&&\hat{H_1}=\left(\begin{array}{c}H_1^1\\ H_1^2\end{array}\right)\sim (1, 2, -1/2, 0),\quad\;\hat{H_2}=\left(\begin{array}{c}H_2^1\\ H_2^2\end{array}\right)\sim(1, 2, 1/2, 0),\nonumber\\
&&\hat{Q}_i=\left(\begin{array}{c}\hat u_i\\ \hat d_i\end{array}\right)\sim(3, 2, 1/6, 1/6), \quad\;\hat{L}_i=\left(\begin{array}{c}\hat \nu_i\\ \hat e_i\end{array}\right)\sim(1, 2, -1/2, -1/2),\nonumber\\
&&\hat{U}^c_i\sim(3, 1, -2/3, -1/6),\quad\; \hat{D}^c_i\sim(3, 1, 1/3, -1/6), \quad\;\hat{E}^c_i\sim(1, 1, 1, 1/2).
\end{eqnarray}

Then, the superpotential in the model can be given by
\begin{eqnarray}
&&W=Y_{u,ij}\hat{Q_i}\hat{H_2}\hat{U_j^c}+\mu \hat{H_1} \hat{H_2}-Y_{d,ij} \hat{Q_i} \hat{H_1} \hat{D_j^c}
-Y_{e,ij} \hat{L_i} \hat{H_1} \hat{E_j^c}\nonumber\\
&&\;\;\;\;\;\;\;\;\;+Y_{\nu, ij}\hat{L_i}\hat{H_2}\hat{\nu}^c_j-\mu' \hat{\eta}_1 \hat{\eta}_2
+Y_{x, ij} \hat{\nu}_i^c \hat{\eta}_1 \hat{\nu}_j^c,
\end{eqnarray}
where $\hat H_1^T$, $\hat H_2^T$, $\hat Q_i^T $, and $\hat L_i^T$ are SU(2) doublet superfields. Note that $\hat U_i^c $, $ \hat D_i^c$, and $\hat E_i^c $ represent up-type quarks, down-type quarks, and charged lepton singlet superfields, respectively. The dimensionless Yukawa coupling parameter $Y$ is a 3$\times$3 matrix. Note that $i,j=1,2,3$ are the generation indices. The summation convention is implied on repeated indices.

Correspondingly, the soft breaking terms of the B-LSSM are generally given as
\begin{eqnarray}
&&\mathcal{L}_{soft}=-
m_{\tilde{q},ij}^2\tilde{Q}_i^*\tilde{Q}_j-m_{\tilde{u},ij}^2(\tilde{u}_i^c)^*\tilde{u}_j^c
-m_{\tilde{d},ij}^2(\tilde{d}_i^c)^*\tilde{d}_j^c-m_{\tilde{L},ij}^2\tilde{L}_i^*\tilde{L}_j
-m_{\tilde{e},ij}^2(\tilde{e}_i^c)^*\tilde{e}_j^c\nonumber\\
&&\hspace{1.4cm}
-m_{\tilde{\nu},ij}^2(\tilde{\nu}_i^c)^* \tilde{\nu}_j^c-m_{\tilde{\eta}_1}^2 |\tilde{\eta}_1|^2-m_{\tilde{\eta}_2}^2 |\tilde{\eta}_2|^2
-m_{H_1}^2|H_1|^2-m_{H_2}^2|H_2|^2\nonumber\\
&&\hspace{1.4cm}
+\Big[-B_\mu H_1H_2 -B_{\mu^{'}}\tilde{\eta}_1 \tilde{\eta}_2 +T_{u,ij}\tilde{Q}_i\tilde{u}_j^cH_2+T_{d,ij}\tilde{Q}_i\tilde{d}_j^cH_1+T_{e,ij}\tilde{L}_i\tilde{e}_j^cH_1\nonumber\\
&&\hspace{1.4cm}
+T_{\nu}^{ij} H_2 \tilde{\nu}_i^c \tilde{L}_j+T_x^{ij} \tilde{\eta}_1 \tilde{\nu}_i^c \tilde{\nu}_j^c-\frac{1}{2}(M_1\tilde{\lambda}_{B} \tilde{\lambda}_{B}+M_2\tilde{\lambda}_{W} \tilde{\lambda}_{W}+M_3\tilde{\lambda}_{g} \tilde{\lambda}_{g}\nonumber\\
&&\hspace{1.4cm}
+2M_{BB^{'}}\tilde{\lambda}_{B^{'}}\tilde{\lambda}_{B}+M_{B^{'}}\tilde{\lambda}_{B^{'}} \tilde{\lambda}_{B^{'}})+h.c.\Big].
\end{eqnarray}
The $SU(2)_L\otimes U(1)_Y\otimes U(1)_{B-L}$ gauge groups break to  $U(1)_{em}$ as the Higgs fields receive vacuum expectation values (VEVs),
\begin{eqnarray}
&&H_1^1=\frac{1}{\sqrt2}(v_1+{\rm Re}H_1^1+i{\rm Im}H_1^1),
\qquad\; H_2^2=\frac{1}{\sqrt2}(v_2+{\rm Re}H_2^2+i{\rm Im}H_2^2),\nonumber\\
&&\tilde{\eta}_1=\frac{1}{\sqrt2}(u_1+{\rm Re}\tilde{\eta}_1+i{\rm Im}\tilde{\eta}_1),
\qquad\;\quad\;\tilde{\eta}_2=\frac{1}{\sqrt2}(u_2+{\rm Re}\tilde{\eta}_2+i{\rm Im}\tilde{\eta}_2)\;.
\end{eqnarray}
Here, $u^2=u_1^2+u_2^2,\; v^2=v_1^2+v_2^2$, $\tan\beta^{'}=\frac{u_2}{u_1}$.

In addition, it is important to consider gauge kinetic mixing, and here we give its covariant derivatives of the form:
\begin{eqnarray}
D_{\mu}=\partial_\mu-iK^TGA,
\end{eqnarray}
where $ K^T = \Big( {Y ,B-L} \Big)$, $ A^T =\Big({A_\mu^{'Y},A_\mu^{'BL}} \Big)$, $K$ is a vector that contains $Y$ and $B-L$ corresponding to hypercharge and $B-L$ charge, and $A_\mu^{'Y}$ and $A_\mu^{'BL}$ are the gauge fields. Note that $G$ is the gauge coupling matrix given as follows:
\begin{eqnarray}
G=\left(\begin{array}{cc}g_{_Y},&g_{_{YB}}^{'}\\g_{_{BY}}^{'},&g_{_{B-L}}\end{array}\right).
\end{eqnarray}
As long as the two Abelian gauge groups are unbroken, one can have the freedom to perform a change of basis by suitable rotation, and $R$ is the proper way to do it:
\begin{eqnarray}
&&\left(\begin{array}{cc}g_{_Y},&g_{_{YB}}^{'}\\g_{_{BY}}^{'},&g_{_{B-L}}\end{array}\right)
R^T=\left(\begin{array}{cc}g_{_1},&g_{_{YB}}\\0,&g_{_{B}}\end{array}\right)\;.
\end{eqnarray}
Here $g_1$ corresponds to the measured hypercharge coupling, which is modified in the B-LSSM and given together with $g_B$ and $g_{YB}$ \cite{BLSSM1}. Next, one can redefine the $U(1)$ gauge fields through
\begin{eqnarray}
&&R\left(\begin{array}{c}A_{_\mu}^{\prime Y} \\ A_{_\mu}^{\prime BL}\end{array}\right)
=\left(\begin{array}{c}A_{_\mu}^{Y} \\ A_{_\mu}^{BL}\end{array}\right)\;.
\end{eqnarray}

An immediate interesting consequence of the gauge kinetic mixing arises in various sectors of the model as discussed in the subsequent analysis. First, the $A^{BL}$ boson mixes at the tree level with the $A^Y$ and $V^3$ bosons. In the basis $(A^Y, V^3, A^{BL})$, the corresponding mass matrix reads
\begin{eqnarray}
&&\left(\begin{array}{*{20}{c}}
\frac{1}{8}g_{_1}^2 v^2 & -\frac{1}{8}g_{_1}g_{_2} v^2 & \frac{1}{8}g_{_1}g_{_{YB}} v^2 \\ [6pt]
-\frac{1}{8}g_{_1}g_{_2} v^2 & \frac{1}{8}g_{_2}^2 v^2 & -\frac{1}{8}g_{_2}g_{_{YB}} v^2\\ [6pt]
\frac{1}{8}g_{_1}g_{_{YB}} v^2 & -\frac{1}{8}g_{_2}g_{_{YB}} v^2 & \frac{1}{8}g_{_{YB}}^2 v^2+\frac{1}{8}g_{_{B}}^2 u^2
\end{array}\right).\label{gauge matrix}
\end{eqnarray}
This mass matrix can be diagonalized by a unitary mixing matrix, which can be expressed by two mixing angles $\theta_{_W}$ and $\theta_{_W}'$ as
\begin{eqnarray}
&&\left(\begin{array}{*{20}{c}}
\gamma\\ [6pt]
Z\\ [6pt]
Z'
\end{array}\right)=
\left(\begin{array}{*{20}{c}}
\cos\theta_{_W} & \sin\theta_{_W} & 0 \\ [6pt]
-\sin\theta_{_W}\cos\theta_{_W}' & \cos\theta_{_W}\cos\theta_{_W}' & \sin\theta_{_W}'\\ [6pt]
\sin\theta_{_W}\sin\theta_{_W}' & -\cos\theta_{_W}'\sin\theta_{_W}' & \cos\theta_{_W}'
\end{array}\right)
\left(\begin{array}{*{20}{c}}
A^Y\\ [6pt]
V^3\\ [6pt]
A^{BL}
\end{array}\right).
\end{eqnarray}
Then $\sin^2\theta_{_W}'$ can be written as
\begin{eqnarray}
\sin^2\theta_{_W}'=\frac{1}{2}-\frac{(g_{_{YB}}^2-g_{_1}^2-g_{_2}^2)x^2+
4g_{_B}^2}{2\sqrt{(g_{_{YB}}^2+g_{_1}^2+g_{_2}^2)x^4+8g_{_B}^2(g_{_{YB}}^2-g_{_1}^2-g_{_2}^2x^2)+16g_{_B}^2}},
\end{eqnarray}
where $x=\frac{v}{u}$. The exact eigenvalues of Eq.(\ref{gauge matrix}) are given by
\begin{eqnarray}
&&\qquad\;\quad\;m_\gamma^2=0,\nonumber\\
&&\qquad\;\quad\;m_{Z,{Z^{'}}}^2=\frac{1}{8}\Big((g_{_1}^2+g_2^2+g_{_{YB}}^2)v^2+4g_{_B}^2u^2 \nonumber\\
&&\qquad\;\qquad\;\qquad\;\mp\sqrt{(g_{_1}^2+g_{_2}^2+g_{_{YB}}^2)^2v^4+8(g_{_{YB}}^2-g_{_1}^2-
g_{_2}^2)g_{_B}^2v^2u^2+16g_{_B}^4u^4}\Big).
\end{eqnarray}

\section{$\mu-e$ conversion in nuclei within the B-LSSM\label{sec3}}
\begin{figure}[htbp]
\setlength{\unitlength}{1mm}
\centering
\includegraphics[width=3in]{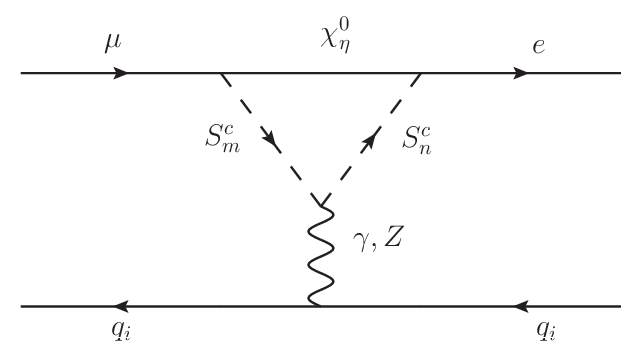}
\caption[]{Penguin-type diagrams for the $\mu-e$ conversion processes at the quark level, where the contributions come from neutral fermion $\chi_\eta^0$ and charged scalar $S_{m,n}^c$ loops.}
\label{fig1}
\end{figure}

\begin{figure}[htbp]
\setlength{\unitlength}{1mm}
\centering
\includegraphics[width=6in]{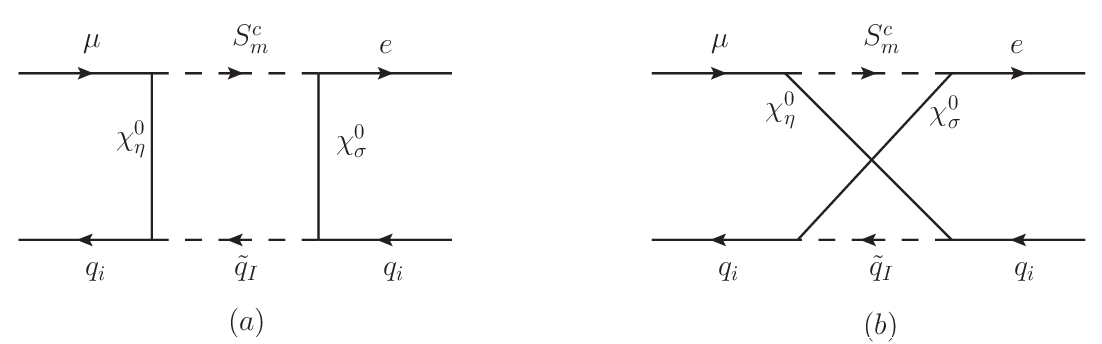}
\caption[]{Box-type diagrams for the $\mu-e$ conversion processes at the quark level, (a) and (b) represent the contributions from neutral fermion $\chi_{\eta,\sigma}^0$, charged scalar $S_m^c$ and squark $\tilde{q}_I$ ($q=u,d$ and $\tilde{u}_I=U_I^+$, $\tilde{d}_I=D_I^-$) loops.}
\label{fig2}
\end{figure}

In this section, we analyze the $\mu-e$ conversion processes at the quark level in the B-LSSM. We give the effective Lagrangian for the $\mu-e$ conversion in nuclei in the following. Both penguin-type diagrams in Fig.~\ref{fig1} and box-type diagrams in Fig.~\ref{fig2} have contributions to the effective Lagrangian. The indices in the figures are $m,n=1,\ldots,6$, $I=1,\ldots,6$, and $\eta,\sigma=1,\ldots,7$.

Figure~\ref{fig1} shows the $\gamma$-penguin-type and $Z$-penguin-type diagrams for the $\mu-e$ conversion processes at the quark level in the B-LSSM. The effective Lagrangian of the $\gamma$-penguin-type diagrams is generally written as
\begin{eqnarray}
&&\mathcal{L}_{int}^{\gamma - {\rm{p}}} = -\frac{{{e^2}}}{{{k^2}}} {\bar e}\Big[{k^2}{\gamma _\alpha }(A_1^L{P_L}+ A_1^R{P_R}) + {m_{\mu}}i{\sigma _{\alpha \beta}}{k^\beta }(A_2^L{P_L}
+ A_2^R{P_R})\Big] \mu  \nonumber\\
&&\qquad\quad\;\; \times \:\sum\limits_{q=u,d} Q_{em}^q{\bar q}{\gamma ^\alpha }q,
\label{eq17}
\end{eqnarray}
where $P_L=(1-\gamma_5)/2$, $P_R=(1+\gamma_5)/2$, $Q^u_{em}=2/3$, $Q^d_{em}=-1/3$, and $m_\mu$ is the muon mass. The coefficients $A_i$ are
\begin{eqnarray}
&&A_1^{L} = \frac{1}{6{m_W^2}}C_R^{S_m^c \chi_\eta^\circ {{\bar l }_i}}C_L^{S_m^{c\ast} {l _j}\bar \chi _\eta ^ \circ }{I_1}({x_{\chi _\eta ^ \circ }},{x_{S_m^c}}),\nonumber\\
&&A_2^{L} = \frac{{{m_{\chi _\eta ^ \circ }}}}{{m_{\mu}}{m_W^2}}C_L^{S_m ^c \chi _\eta ^ \circ {{\bar l }_i}}C_L^{S_m ^{c\ast} {l _j}\bar \chi _\eta ^ \circ }
\Big[ {I_3}({x_{\chi _\eta ^ \circ }},{x_{S_m ^ c }}) - {I_2}(
{x_{\chi _\eta ^ \circ }},{x_{S_m ^c }}) \Big], \nonumber\\
&&A_{1,2}^{R} = \left. {A_{1,2}^{L}} \right|{ _{L \leftrightarrow R}},
\end{eqnarray}
where $x_i=m_i^2/m_W^2$, $I_i$ is the loop function, and $C$ is the coupling which can be found in the appendixes.

The effective Lagrangian of the $Z$-penguin-type diagrams is generally written as
\begin{eqnarray}
&&\mathcal{L}_{int}^{Z - {\rm{p}}} = \frac{{{e^2}}}{{m_Z^2}s_{_W}^2c_{_W}^2}\sum\limits_{q=u,d}\frac{Z_L^q+Z_R^q}{2}{\bar q}\gamma _\alpha q {\bar e}{\gamma ^\alpha }({B_L}{P_L} + {B_R}{P_R})\mu  ,
\end{eqnarray}
where
\begin{eqnarray}
Z_{L,R}^q = T_{3L,R}^q - Q_{em}^q s_{_W}^2,\quad (q=u,d),
\end{eqnarray}
with $T_{3L}^u=\frac{1}{2}$, $T_{3L}^d=-\frac{1}{2}$ and $T_{3R}^u=T_{3R}^d=0$. The contributions to the coefficients $B_{L,R}$ are
\begin{eqnarray}
&&B_L = \, \frac{1}{2{e^2}}C_R^{S_n^ c \chi _\eta ^0{{\bar l }_i}}
C_R^{ZS_m ^ c S_n ^ {c\ast} }C_L^{S_m ^{c\ast} {l _j}\bar \chi _\eta ^0}{G_1}
({x_{\chi _\eta ^0}},{x_{S_m ^ c }},{x_{S_n ^ c }}) ,
\nonumber\\
&&B_R = \left. {B_L} \right|{ _{L \leftrightarrow R}}.
\end{eqnarray}
Here $G_i$ is the loop function which can be found in the appendixes.

The effective Lagrangian of the box-type diagrams shown in Fig.~\ref{fig2} is generally written as
\begin{eqnarray}
&&\mathcal{L}_{int}^{box}  = e^2 \sum\limits_{q=u,d}{\bar q}\gamma _\alpha q {\bar e}{\gamma ^\alpha }({D_q^L}{P_L} + {D_q^R}{P_R})\mu ,
\end{eqnarray}
with
\begin{eqnarray}
&&D_q^L = \frac{1}{8{e^2}{m_W^2}}{G_3}({x_{\chi_\eta^0}},{x_{\chi_\sigma^0}},{x_{S_m^c}},
{x_{\tilde{q}_I}})\Big[
C_R^{S_m ^ {c\ast}l_j{\bar \chi} _\eta^0} C_R^{S_m^ c \chi _\sigma^0{{\bar l }_i}}
C_R^{{\tilde{q}_I} \chi _\eta^0{{\bar q }_i}}C_R^{{\tilde{q}_I} \chi _\sigma^0{{\bar q }_i}\ast}
\nonumber\\
&&\qquad\qquad - \:C_R^{S_m ^ {c\ast}l_j{\bar \chi} _\eta^0} C_R^{S_m^ c \chi _\sigma^0{{\bar l }_i}}C_L^{{\tilde{q}_I} \chi _\eta^0{{\bar q }_i}\ast}C_L^{{\tilde{q}_I} \chi _\sigma^0{{\bar q }_i}}\Big] \nonumber\\
&&\qquad\qquad - \:\frac{{m_{\chi_\eta^0}}{m_{\chi_\sigma^0}}}{4{e^2}{m_W^4}}
{G_2}({x_{\chi_\eta^0}},{x_{\chi_\sigma^0}},{x_{S_m^c}},
{x_{\tilde{q}_I}})\Big[
C_R^{S_m ^ {c\ast}l_j{\bar \chi} _\eta^0} C_R^{S_m^ c \chi _\sigma^0{{\bar l }_i}}
C_L^{{\tilde{q}_I} \chi _\eta^0{{\bar q }_i}}C_L^{{\tilde{q}_I} \chi _\sigma^0{{\bar q }_i}\ast}
\nonumber\\
&&\qquad\qquad - \:C_R^{S_m ^ {c\ast}l_j{\bar \chi} _\eta^0} C_R^{S_m^ c \chi _\sigma^0{{\bar l }_i}}C_R^{{\tilde{q}_I} \chi _\eta^0{{\bar q }_i}\ast}C_R^{{\tilde{q}_I} \chi _\sigma^0{{\bar q }_i}}\Big] ,
\nonumber\\
&&D_q^R = \left. {F_q^L} \right|
{ _{L \leftrightarrow R}} \:,\:\;(q=u,d \;\textrm{and}\; \tilde{u}_I=U_I^+, \:\tilde{d}_I=D_I^-).
\end{eqnarray}

Using the expression for the effective Lagrangian of the $\mu-e$ conversion processes at the quark level, one can calculate the $\mu-e$ conversion rate in a nucleus~\cite{Bernabeu}:
\begin{eqnarray}
&&{\rm{CR}}(\mu \to e:{\rm{Nucleus}}) \nonumber\\
&&\qquad = 4 \alpha^5 \frac{Z_{\rm{eff}}^4}{Z } \left| F(q^2) \right|^2 m_\mu^5  \Big[\left| Z( A_1^L  -  {A_2^R} ) - (2Z+N)\bar{D}_u^L - (Z+2N)\bar{D}_d^L \right| ^2 \nonumber\\
&&\qquad\quad + \: \left| Z( A_1^R  -  {A_2^L} ) - (2Z+N)\bar{D}_u^R - (Z+2N)\bar{D}_d^R \right|^2 \Big]\frac{1}{\Gamma_{\rm{capt}}},
\label{gamma-2}
\end{eqnarray}
with
\begin{eqnarray}
&&\bar{D}_q^L = D_q^L + \frac{Z_L^q+Z_R^q}{2} \frac{B_L}{{m_Z^2}s_{_W}^2c_{_W}^2} , \nonumber\\
&&\bar{D}_q^R  =  \left. {\bar{D}_q^L} \right|{ _{L \leftrightarrow R}}  \quad (q=u,d ), \quad
\end{eqnarray}
where $Z$ is the number of protons in the nucleus and $N$ is the number of neutrons in the nucleus. Note that $Z_{\rm{eff}}$ is an effective atomic charge~\cite{Zeff,Zeff1}, $F(q^2)$ is the nuclear form factor, and $\Gamma_{\rm{capt}}$ is the total muon capture rate. In the following numerical analysis, we consider the $\mu - e$ conversion rate in $_{22}^{48}{\rm{Ti}}$, $_{\:79}^{197}{\rm{Au}}$ and $_{13}^{27}{\rm{Al}}$ nuclei, where the values of $Z_{\rm{eff}}$, $F(q^2\simeq-m_\mu^2)$, and $\Gamma_{(\rm{capt})}$ for the different nuclei can be seen in Table.~\ref{table1} and follow Ref.~\cite{Kitano}.

\begin{table*}
\begin{tabular*}{\textwidth}{@{\extracolsep{\fill}}lllll@{}}
\hline
$_{Z}^{A}{\rm{Nucleus}}$ & \multicolumn{1}{c}{$Z_{\rm{eff}}$} & \multicolumn{1}{c}{$F(q^2\simeq-m_\mu^2)$} & $\Gamma_{\rm{capt}}({\rm{GeV}})$ \\
\hline
$_{22}^{48}{\rm{Ti}}$ & 17.6 & 0.54 & $1.70422\times10^{-18}$ \\
$_{\:79}^{197}{\rm{Au}}$ & 33.5 & 0.16 & $8.59868\times10^{-18}$ \\
$_{13}^{27}{\rm{Al}}$ & 11.5 & 0.64 & $0.464079\times10^{-18}$ \\
\hline
\end{tabular*}
\caption{Values of $Z_{\rm{eff}}$, $F(q^2\simeq-m_\mu^2)$, and $\Gamma_{\rm{capt}}$ for different nuclei.}
\label{table1}
\end{table*}

\section{Numerical analysis\label{sec4}}
The relevant SM input parameters are chosen as $m_W$\emph{=}$80.385~{\rm GeV}$, $m_Z$\emph{=}$90.1876~{\rm GeV}$, $\alpha_{em}(m_Z)=1/128.9$, and $\alpha_s(m_Z)=0.118$. Considering that the updated experimental data  on searching $Z'$ indicate $M_{Z'}\geq 4.05~{\rm TeV}$ at 95\% C.L.~\cite{newZ}, we choose $M_{Z'}$\emph{=}$4.2~{\rm TeV}$ in the following. References~\cite{GCG,MAB} give an upper bound on the ratio between the $Z'$ mass and its gauge coupling at 99\% C.L. as $M_{Z'}/g_B\geq 6~{\rm TeV}$, and then the scope of $g_B$ is $0<g_B<0.7$. LHC experimental data constrain $\tan\beta'<1.5$~\cite{48}. Considering the constraint of the experiments~\cite{PDGPA}, we take $M_{1}$\emph{=}$500~{\rm GeV},\;M_{2}$\emph{=}$600~{\rm GeV}$, $B_\mu'$\emph{=}$5\times10^5~{\rm GeV}^2$, $A_e=0.5$ TeV, $m_{\tilde{q}}$\emph{=}$m_{\tilde{u}}$\emph{=}$m_{\tilde{d}}$\emph{=}$diag(2, 2, 1.6)~{\rm TeV}$, $T_u$\emph{=}$Y_u\times$$diag(1, 1, 1)~{\rm TeV}$, $T_d$\emph{=}$Y_d\times$$diag(1, 1, 1)~{\rm TeV}$, and $T_x$\emph{=}$diag(1, 1, 1)~{\rm TeV}$, respectively.

We need to consider the constraint of the SM-like Higgs boson mass. The remaining key parameters that affect the Higgs boson mass are $\tan\beta$, $\tan\beta'$, $g_B$, and $g_{YB}$. By constantly adjusting the parameters, the final numerical analysis strictly conforms to the constraint of the SM-like Higgs boson measured mass  $m_h$\emph{=}$125.09\pm0.24~{\rm GeV}$ in 3$\sigma$~\cite{PDGPA}. In addition, it should be noted that although the B-LSSM can produce nonzero neutrinos, the mass of the neutrinos is too small to affect the problem we study, so we approximately consider the mass of neutrinos to be zero. Although the B-LSSM contains LFV sources in the neutrino Yukawa sector, such as the $Y_\nu$ matrix, the neutrino oscillation causes $Y_\nu \sim \mathcal{O}(10^{-6})$, which contributes very little to the problem we study; thus we approximately ignore the influence of the neutrino Yukawa sector in the numerical analysis.

Since we are studying the lepton flavor violating processes, we have to consider the off-diagonal terms for the soft breaking slepton mass matrices $m^2_{\bar L, \bar e}$ and the trilinear coupling matrix $T_e$, which are defined by~\cite{sl-mix,sl-mix1,sl-mix2,sl-mix3,sl-mix4,neu-zhang2}
\begin{eqnarray}
&&\hspace{-0.75cm}\quad\;\,{m^2_{\tilde L}} = \left( {\begin{array}{*{20}{c}}
   1 & \delta_{12}^{LL} & \delta_{13}^{LL}  \\
   \delta_{12}^{LL} & 1 & \delta_{23}^{LL}  \\
   \delta_{13}^{LL} & \delta_{23}^{LL} & 1  \\
\end{array}} \right){m_L^2},\\
&&\hspace{-0.75cm}\quad\:{m_{\tilde e^c}^2} = \left( {\begin{array}{*{20}{c}}
   1 & \delta_{12}^{RR} & \delta_{13}^{RR}  \\
   \delta_{12}^{RR} & 1 & \delta_{23}^{RR}  \\
   \delta_{13}^{RR} & \delta_{23}^{RR} & 1  \\
\end{array}} \right){m_E^2},\\
&&T_e=
\left(\begin{array}{ccc} 1& \delta_{12}^{LR} &\delta_{13}^{LR}
\\ \delta_{12}^{LR} &1 &\delta_{23}^{LR}
\\ \delta_{13}^{LR} &\delta_{23}^{LR}  &1\end{array}\right){A_e}.
\end{eqnarray}
We know that LFV processes are flavor dependent, just as the LFV rate for $\mu - e$ transitions depends on the slepton mixing parameters $\delta_{12}^{XX}~(X=L,R)$; thus we only need to consider the effect of slepton mixing parameters  $\delta_{12}^{XX}~(X=L,R)$ on the $\mu-e$ conversion rate. The other slepton mixing parameters  $\delta_{13}^{XX}~(X=L,R)$ and $\delta_{23}^{XX}~(X=L,R)$ have no effect on the $\mu-e$ conversion rate, so we choose $\delta_{13}^{XX}~(X=L,R)=0$ and $\delta_{23}^{XX}~(X=L,R)=0$.

In the subsequent numerical analysis, we not only give different sensitive parameters on the effect of the $\mu-e$ conversion rate in nuclei, but we also give to the influence of parameters on ${\rm{Br}}(\mu \rightarrow e\gamma)$ in the B-LSSM.
Constrained by the ${\rm{Br}}(\mu \rightarrow e\gamma)$, the final magnitude of the $\mu-e$ conversion rate can be achieved.

\subsection{Effect of slepton mixing parameters  $\delta_{12}^{XX}~(X=L,R)$ on the $\mu-e$ conversion rate }
In this subsection, we plot the influence of slepton mixing parameters  $\delta_{12}^{XX}~(X=L,R)$ on the $\mu - e$ conversion rate and ${\rm{Br}}(\mu \rightarrow e\gamma)$. Here, we choose $\tan\beta=11$, $\tan\beta'=1.3$, $g_B=0.5$, $g_{YB}=-0.4$, and $m_L=m_E=1$ TeV as fixed values to study the influence of $\delta_{12}^{XX}~(X=L,R)$ on the $\mu-e$ conversion rate in different nuclei. When the variable is $\delta_{12}^{XX}~(X=L,R)$ in the figures, the other two are $\delta_{12}^{XX}~(X=L,R)=0$. In the figures, the dashed and dot-dashed lines denote the present limits and future sensitivities respectively; the red solid line is ruled out by the present limit of ${\rm{Br}}(\mu \rightarrow e\gamma)$, and the black solid line is consistent with the present limit of ${\rm{Br}}(\mu \rightarrow e\gamma)$.

\begin{figure}
\setlength{\unitlength}{1mm}
\centering
\includegraphics[width=2.8in]{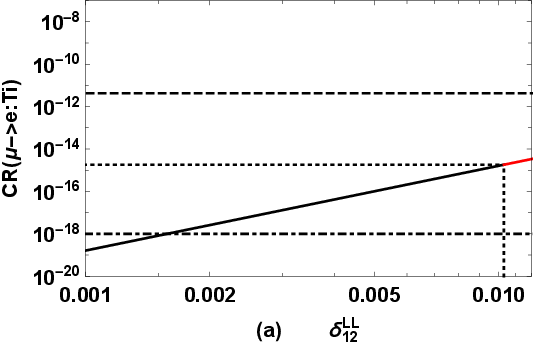}
\vspace{0cm}
\includegraphics[width=2.8in]{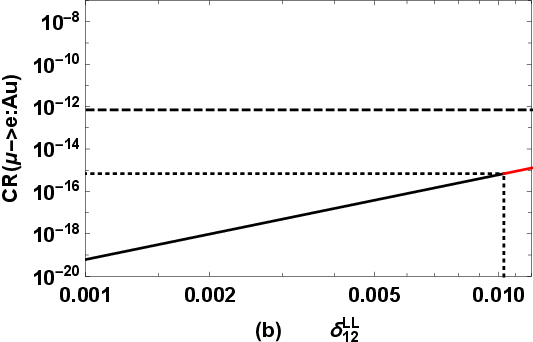}
\vspace{0cm}
\includegraphics[width=2.8in]{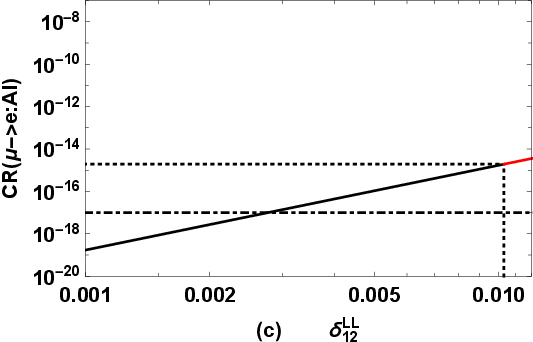}
\vspace{0cm}
\includegraphics[width=2.8in]{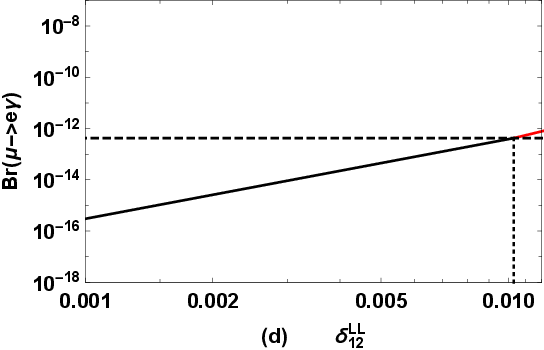}
\vspace{0cm}
\caption[]{{\label{3}} (a) ${\rm{CR}}(\mu\rightarrow e:\rm{Ti})$, (b) ${\rm{CR}}(\mu\rightarrow e:\rm{Au})$, and (c) ${\rm{CR}}(\mu\rightarrow e:\rm{Al})$ versus the slepton flavor mixing parameter $\delta_{12}^{LL}$, where the dashed lines in panels (a) and (b) stand for the upper limits on ${\rm{CR}}(\mu\rightarrow e:\rm{Ti})$ and ${\rm{CR}}(\mu\rightarrow e:\rm{Au})$, respectively, and the dot-dashed line in panels (a) and (c) represents the sensitivity of future experiments on ${\rm{CR}}(\mu\rightarrow e:\rm{Ti})$ and ${\rm{CR}}(\mu\rightarrow e:\rm{Al})$, respectively. (d) ${\rm{Br}}(\mu \rightarrow e\gamma)$ versus slepton mixing parameter $\delta_{12}^{LL}$, where the dashed line denotes the present limit of ${\rm{Br}}(\mu\rightarrow e\gamma)$ at 90\% C.L. as shown in Eq.(\ref{a4}). Here, the red solid line is ruled out by the present limit of ${\rm{Br}}(\mu \rightarrow e\gamma)$, and the black solid line is consistent with the present limit of ${\rm{Br}}(\mu \rightarrow e\gamma)$.}
\end{figure}

\begin{figure}
\setlength{\unitlength}{1mm}
\centering
\includegraphics[width=2.8in]{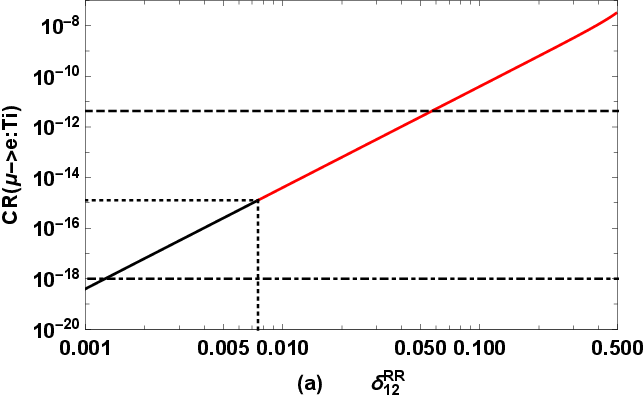}
\vspace{0cm}
\includegraphics[width=2.8in]{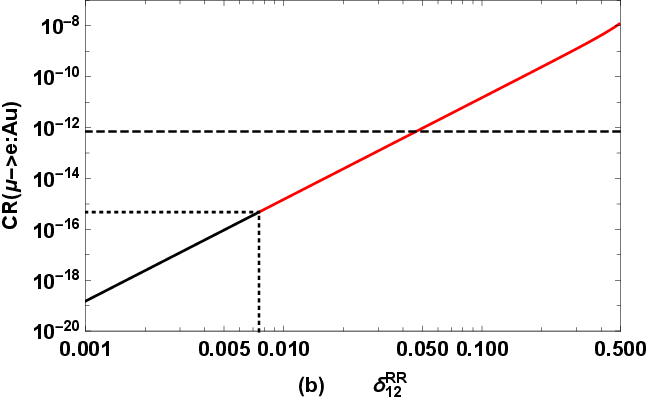}
\vspace{0cm}
\includegraphics[width=2.8in]{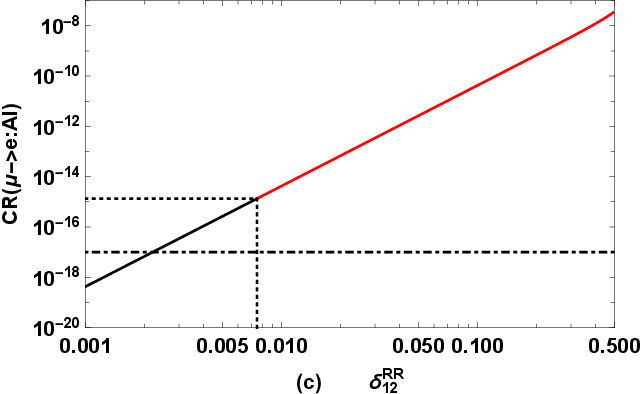}
\vspace{0cm}
\includegraphics[width=2.8in]{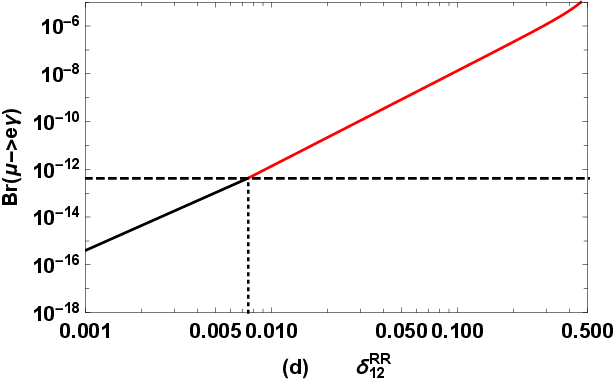}
\vspace{0cm}
\caption[]{{\label{4}} (a) ${\rm{CR}}(\mu\rightarrow e:\rm{Ti})$, (b) ${\rm{CR}}(\mu\rightarrow e:\rm{Au})$, (c) ${\rm{CR}}(\mu\rightarrow e:\rm{Al})$, and (d) ${\rm{Br}}(\mu \rightarrow e\gamma)$ versus the slepton flavor mixing parameter $\delta_{12}^{RR}$.}
\end{figure}

\begin{figure}
\setlength{\unitlength}{1mm}
\centering
\includegraphics[width=2.8in]{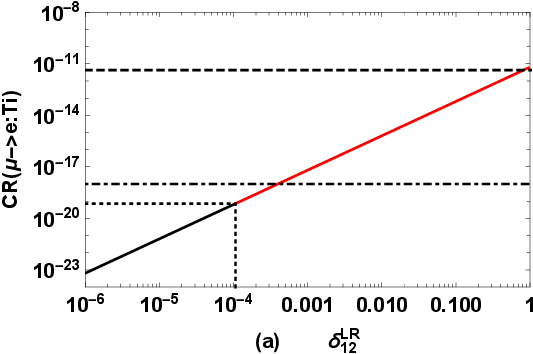}
\vspace{0cm}
\includegraphics[width=2.8in]{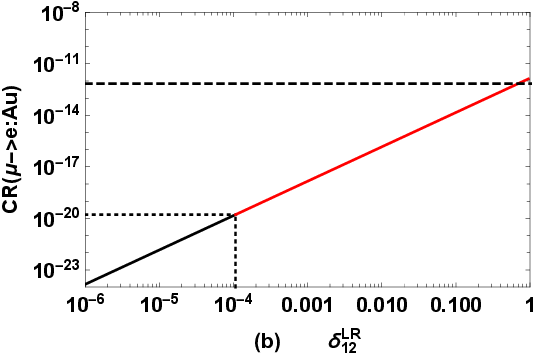}
\vspace{0cm}
\includegraphics[width=2.8in]{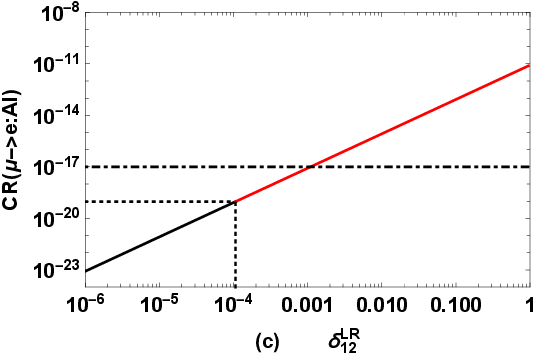}
\vspace{0cm}
\includegraphics[width=2.8in]{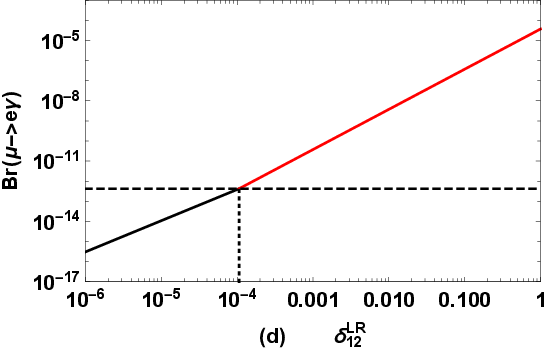}
\vspace{0cm}
\caption[]{{\label{5}} (a) ${\rm{CR}}(\mu\rightarrow e:\rm{Ti})$, (b) ${\rm{CR}}(\mu\rightarrow e:\rm{Au})$, (c) ${\rm{CR}}(\mu\rightarrow e:\rm{Al})$, and (d) ${\rm{Br}}(\mu \rightarrow e\gamma)$ versus the slepton flavor mixing parameter $\delta_{12}^{LR}$.}
\end{figure}

In Fig.~\ref{3}, we plot the $\mu - e$ conversion rate in the different nuclei and ${\rm{Br}}(\mu\rightarrow e\gamma)$ versus $\delta_{12}^{LL}$ for $\delta_{12}^{RR}=\delta_{12}^{LR}=0$. It is obvious that LFV rates increase with the increase of the slepton flavor mixing parameter $\delta_{12}^{LL}$ because the LFV processes are flavor dependent, and the LFV rate for $\mu - e$ transitions depends on the slepton mixing parameters $\delta_{12}^{XX}~(X=L,R)$. It can be seen from the figure that ${\rm{Br}}(\mu \rightarrow e\gamma)$ can reach the experimental upper limit, but the $\mu - e$ conversion rate in the nuclei cannot. When we consider the constraint of ${\rm{Br}}(\mu\rightarrow e\gamma)$ to the $\mu - e$ conversion rate, from Figs.~\ref{3}(a) and (c), ${\rm{CR}}(\mu\rightarrow e:\rm{Ti})$ and ${\rm{CR}}(\mu\rightarrow e:\rm{Al})$ can exceed $\mathcal{O}(10^{-15})$ and above their respective future experimental sensitivities. Thus, there is still hope that the high future experimental sensitivities will detect ${\rm{CR}}(\mu\rightarrow e:\rm{Ti})$ and ${\rm{CR}}(\mu\rightarrow e:\rm{Al})$. From Fig.~\ref{3}(b) ${\rm{CR}}(\mu\rightarrow e:\rm{Au})$ can exceed $\mathcal{O}(10^{-16})$. Therefore, it can be seen that the limit of ${\rm{Br}}(\mu\rightarrow e\gamma)$ to the $\mu - e$ conversion rate in the nuclei is very strict.

Figures~\ref{4}(a)-~\ref{4}(c) represent the relationship of ${\rm{CR}}(\mu\rightarrow e:\rm{Ti})$ ,${\rm{CR}}(\mu\rightarrow e:\rm{Au})$, and ${\rm{CR}}(\mu\rightarrow e:\rm{Al})$  with changes of the slepton flavor mixing parameter $\delta_{12}^{RR}$, respectively, and Fig~\ref{4}(d) represents the relationship of ${\rm{Br}}(\mu \rightarrow e\gamma)$ with changes of $\delta_{12}^{RR}$. The general trend shown in the four graphs is that as $\delta_{12}^{RR}$ continues to increase, ${\rm{CR}}(\mu\rightarrow e:\rm{Ti})$, ${\rm{CR}}(\mu\rightarrow e:\rm{Au})$, ${\rm{CR}}(\mu\rightarrow e:\rm{Al})$ and ${\rm{Br}}(\mu \rightarrow e\gamma)$ also increase. As can be seen from Fig.~\ref{4}(a),~\ref{4}(b) and ~\ref{4}(d), ${\rm{CR}}(\mu\rightarrow e:\rm{Ti})$, ${\rm{CR}}(\mu\rightarrow e:\rm{Au})$, and ${\rm{Br}}(\mu \rightarrow e\gamma)$ can all exceed their respective present experimental limits, but due to the restriction of the limit of ${\rm{Br}}(\mu \rightarrow e\gamma)$, ${\rm{CR}}(\mu\rightarrow e:\rm{Ti})$ and ${\rm{CR}}(\mu\rightarrow e:\rm{Au})$ can only be far below the experimental limits. For Fig.~\ref{4}(a) and ~\ref{4}(c), ${\rm{CR}}(\mu\rightarrow e:\rm{Ti})$ and ${\rm{CR}}(\mu\rightarrow e:\rm{Al})$ can exceed the sensitivity of future experiments under the limit of ${\rm{Br}}(\mu \rightarrow e\gamma)$.

Because the LFV processes are flavor dependent, $\delta_{12}^{LR}$ also has a greater influence on the $\mu - e$ conversion rate in the different nuclei and ${\rm{Br}}(\mu \rightarrow e\gamma)$. As $\delta_{12}^{LR}$ increases, ${\rm{CR}}(\mu\rightarrow e:\rm{Ti})$, ${\rm{CR}}(\mu\rightarrow e:\rm{Au})$, ${\rm{CR}}(\mu\rightarrow e:\rm{Al})$ and ${\rm{Br}}(\mu \rightarrow e\gamma)$ also increase. As seen from Fig.~\ref{5}, ${\rm{Br}}(\mu \rightarrow e\gamma)$ can quickly exceed the experimental limit. Figure~\ref{5}(d) shows that the present experimental limit bound of ${\rm{Br}}(\mu \rightarrow e\gamma)$ constrains $\delta_{12}^{LR}<10^{-4}$. Considering the constraint of ${\rm{Br}}(\mu \rightarrow e\gamma)$ to the $\mu - e$ conversion rate in different nuclei, ${\rm{CR}}(\mu\rightarrow e:\rm{Ti})$ and  ${\rm{CR}}(\mu\rightarrow e:\rm{Au})$ cannot reach their respective current experimental upper limits, and neither ${\rm{CR}}(\mu\rightarrow e:\rm{Ti})$ nor ${\rm{CR}}(\mu\rightarrow e:\rm{Al})$ can reach the sensitivity of their respective future experiments, which indicates that the constraint of ${\rm{Br}}(\mu \rightarrow e\gamma)$ to the $\mu - e$ conversion rate in different nuclei is very obvious for the slepton flavor mixing parameter $\delta_{12}^{LR}$.

\subsection{Effect of $M_E$, $\tan\beta'$, and $g_{YB}$ on the $\mu-e$ conversion rate }
In this section, we study the influence of other basic parameters on the $\mu-e$ conversion rate and ${\rm{Br}}(\mu \rightarrow e\gamma)$. We first set appropriate numerical values for slepton flavor mixing parameters, such as $\delta_{12}^{LL}=0.01$, $\delta_{12}^{RR}=0.006$, and $\delta_{12}^{LR}=1\times10^{-4}$. We also keep neutral fermion masses $m_{\chi_\eta^o}>200~{\rm GeV}$ $(\eta=1,\cdots,7)$, the scalar masses $m_{S_{m, n}^{c}}>500~{\rm GeV}$ $(m,n=1,\cdots,6)$ and the SM-like Higgs boson mass $m_h$\emph{=}$125.09\pm 0.24\: {\rm{GeV}}$ in 3$\sigma$ to avoid the range ruled out by the experiments. Then we research the influence of the basic parameters $m_L=m_E\equiv M_E$, $\tan\beta'$, and $g_{YB}$ on the $\mu-e$ conversion rate and ${\rm{Br}}( \mu\rightarrow e\gamma)$, respectively.

\begin{figure}
\setlength{\unitlength}{1mm}
\centering
\includegraphics[width=2.8in]{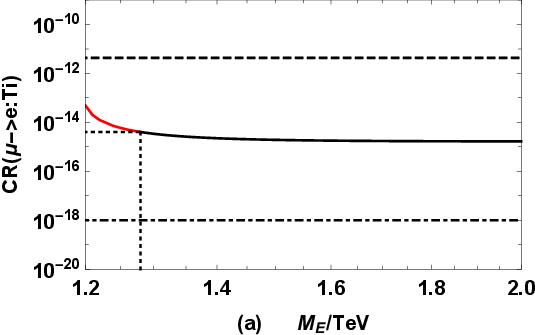}
\vspace{0cm}
\includegraphics[width=2.8in]{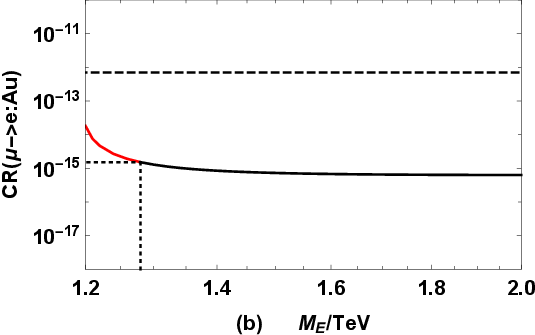}
\vspace{0cm}
\includegraphics[width=2.8in]{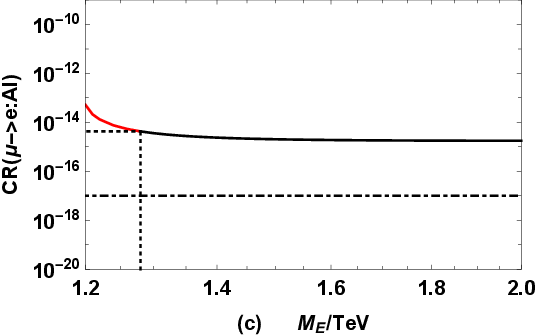}
\vspace{0cm}
\includegraphics[width=2.8in]{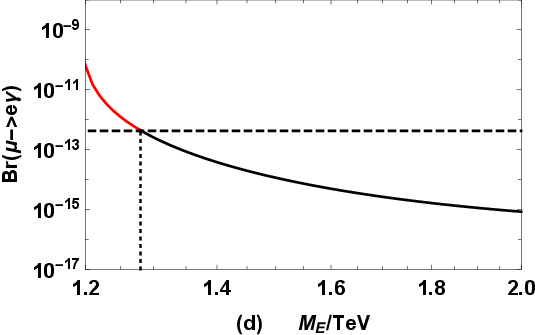}
\vspace{0cm}
\caption[]{{\label{6}} (a) ${\rm{CR}}(\mu\rightarrow e:\rm{Ti})$, (b) ${\rm{CR}}(\mu\rightarrow e:\rm{Au})$, and (c) ${\rm{CR}}(\mu\rightarrow e:\rm{Al})$ versus parameters $M_E$, where the dashed lines in panels (a) and (b) stand for the upper limits on ${\rm{CR}}(\mu\rightarrow e:\rm{Ti})$ and ${\rm{CR}}(\mu\rightarrow e:\rm{Au})$, respectively, and the dot-dashed lines in panels (a) and (c) represent the sensitivity of future experiments on ${\rm{CR}}(\mu\rightarrow e:\rm{Ti})$ and ${\rm{CR}}(\mu\rightarrow e:\rm{Al})$, respectively. (d) ${\rm{Br}}(\mu \rightarrow e\gamma)$ versus parameters $M_E$, where the dashed line denotes the present limit of ${\rm{Br}}(\mu\rightarrow e\gamma)$ at 90\% C.L. as shown in Eq.(\ref{a4}). Here, the red solid line is ruled out by the present limit of ${\rm{Br}}(\mu \rightarrow e\gamma)$, and the black solid line is consistent with the present limit of ${\rm{Br}}(\mu \rightarrow e\gamma)$.}
\end{figure}

\begin{figure}
\setlength{\unitlength}{1mm}
\centering
\includegraphics[width=2.8in]{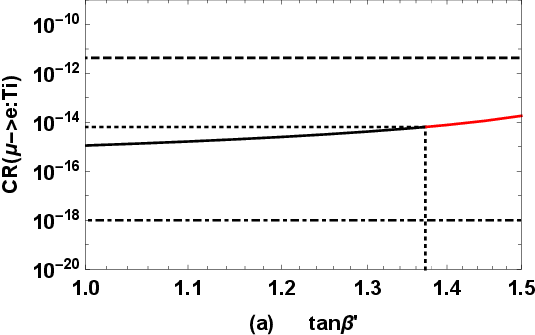}
\vspace{0cm}
\includegraphics[width=2.8in]{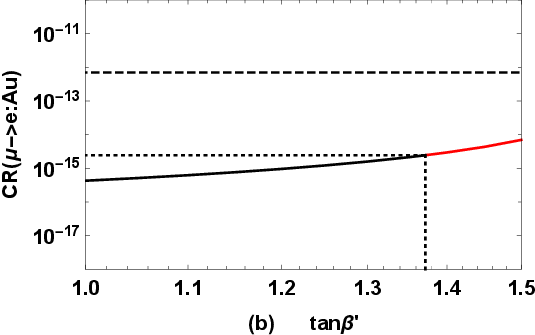}
\vspace{0cm}
\includegraphics[width=2.8in]{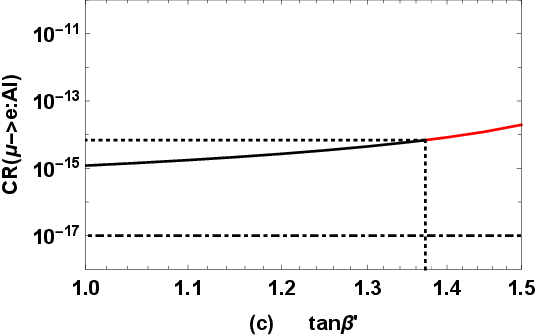}
\vspace{0cm}
\includegraphics[width=2.8in]{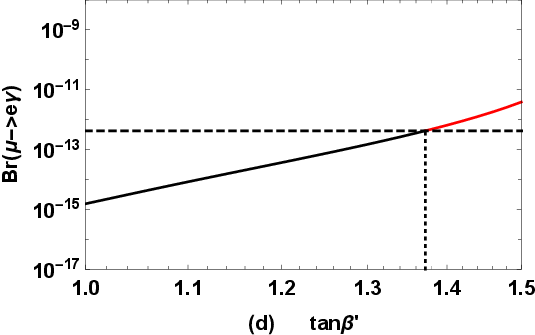}
\vspace{0cm}
\caption[]{{\label{7}} (a) ${\rm{CR}}(\mu\rightarrow e:\rm{Ti})$, (b) ${\rm{CR}}(\mu\rightarrow e:\rm{Au})$, (c) ${\rm{CR}}(\mu\rightarrow e:\rm{Al})$, and (d) ${\rm{Br}}(\mu \rightarrow e\gamma)$ versus parameters $\tan\beta'$.}
\end{figure}
\begin{figure}
\setlength{\unitlength}{1mm}
\centering
\includegraphics[width=2.8in]{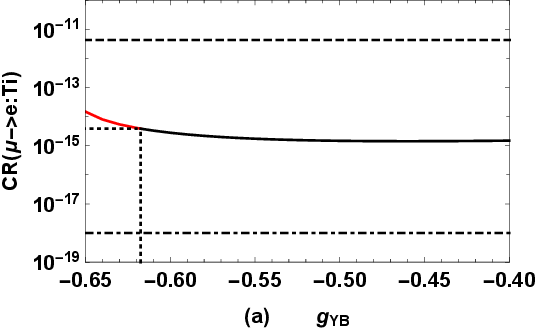}
\vspace{0cm}
\includegraphics[width=2.8in]{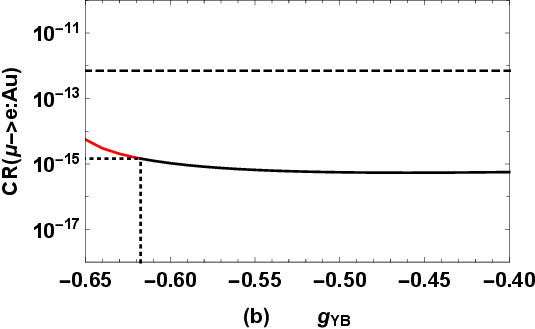}
\vspace{0cm}
\includegraphics[width=2.8in]{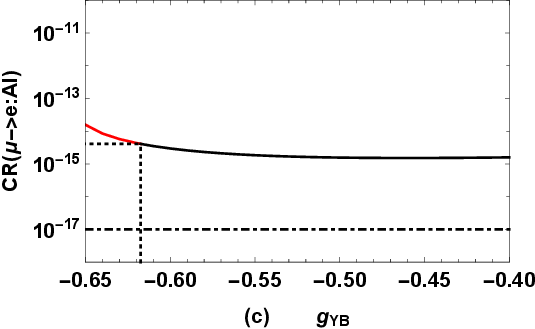}
\vspace{0cm}
\includegraphics[width=2.8in]{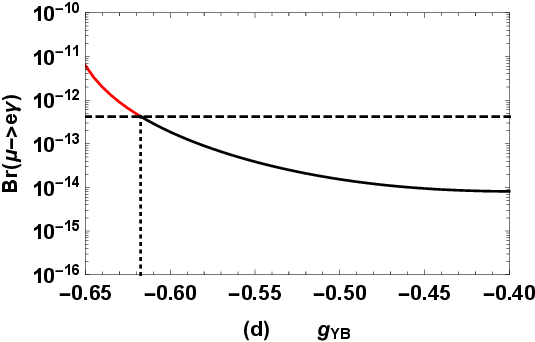}
\vspace{0cm}
\caption[]{{\label{8}} (a) ${\rm{CR}}(\mu\rightarrow e:\rm{Ti})$, (b) ${\rm{CR}}(\mu\rightarrow e:\rm{Au})$, (c) ${\rm{CR}}(\mu\rightarrow e:\rm{Al})$, and (d) ${\rm{Br}}(\mu \rightarrow e\gamma)$ versus parameters $g_{YB}$.}
\end{figure}

We plot the $\mu - e$ conversion rates in the different nuclei and ${\rm{Br}}(\mu\rightarrow e\gamma)$ versus $M_E$ in Fig.~\ref{6}. When we study the influence of $M_E$ on the $\mu - e$ conversion rate and ${\rm{Br}}(\mu\rightarrow e\gamma)$, the values of other basic parameters are $\tan\beta=10$, $\tan\beta'=1.4$, $g_B=0.5$ and $g_{YB}=-0.5$ respectively. In Fig.~\ref{6}, it is obvious that the $\mu - e$ conversion rate and ${\rm{Br}}(\mu\rightarrow e\gamma)$ decrease with the increase of $M_E$, due to the fact that the mass of sleptons increases as $M_E$ increases, which indicates that heavy sleptons play a suppressive role in the rates of LFV processes. Although ${\rm{CR}}(\mu\rightarrow e:\rm{Ti})$ and ${\rm{CR}}(\mu\rightarrow e:\rm{Au})$ cannot reach their current upper limit under the constraint of ${\rm{Br}}(\mu \rightarrow e\gamma)$, ${\rm{CR}}(\mu\rightarrow e:\rm{Ti})$ and ${\rm{CR}}(\mu\rightarrow e:\rm{Al})$ can reach the high future experimental sensitivities with small $M_E$.

In order to see the effect of $\tan\beta'$, which includes new parameters in the B-LSSM beyond the MSSM, we plot the $\mu - e$ conversion rate in the different nuclei and ${\rm{Br}}(\mu\rightarrow e\gamma)$ versus $\tan\beta'$ in Fig.~\ref{7}, choosing the values of the other basic parameters as $\tan\beta=11.2$, $g_B=0.2$, $g_{YB}=-0.1$, and $m_L=m_E=1$ TeV. Figure~\ref{7} shows that LFV rates increase with the increasing of $\tan\beta'$. ${\rm{Br}}(\mu\rightarrow e\gamma)$ can reach the experimental upper limit but ${\rm{CR}}(\mu\rightarrow e:\rm{Ti})$ and ${\rm{CR}}(\mu\rightarrow e:\rm{Au})$ cannot. When $\tan\beta'\leq1.37$ under the constraint of ${\rm{Br}}(\mu\rightarrow e\gamma)$, although ${\rm{CR}}(\mu\rightarrow e:\rm{Ti})$ and ${\rm{CR}}(\mu\rightarrow e:\rm{Au})$ cannot reach their current upper limits, ${\rm{CR}}(\mu\rightarrow e:\rm{Ti})$ and ${\rm{CR}}(\mu\rightarrow e:\rm{Al})$ can exceed the sensitivity of future experiments. This shows that in the near future, with the continuous improvement of experimental accuracy, the $\mu - e$ conversion rate in the different nuclei can be detected.

In Fig.~\ref{8}, we draw the influence of $g_{YB}$ on the $\mu - e$ conversion rate and ${\rm{Br}}(\mu\rightarrow e\gamma)$, where $g_{YB}$ is also a new parameter in the B-LSSM beyond the MSSM. We choose the values of other basic parameters except $g_{YB}$ as $\tan\beta=20$, $\tan\beta'=1.15$, $g_B=0.5$ and $m_L=m_E=1$ TeV  respectively. It can be seen from Fig.~\ref{8} that the $\mu - e$ conversion rate and ${\rm{Br}}(\mu\rightarrow e\gamma)$ decrease with the increase of $g_{YB}$. When $g_{YB}$ is small, ${\rm{Br}}(\mu\rightarrow e\gamma)$ can reach the experimental upper limit, but ${\rm{CR}}(\mu\rightarrow e:\rm{Ti})$ and ${\rm{CR}}(\mu\rightarrow e:\rm{Au})$ cannot. Note that $g_{YB}$ affects the numerical results through the new mass matrix of sleptons, Higgs bosons, and neutralinos, which can make contributions to these LFV processes.

\begin{table*}
\begin{tabular*}{\textwidth}{@{\extracolsep{\fill}}llll@{}}
\hline
Parameters&Min&Max\\
\hline
$\tan\beta$&1&50&\\
$g_{YB}$&-0.7&-0.1&\\
$g_{B}$&0.1&0.7&\\
$\tan\beta'$&1&1.5&\\
$M_E/{\rm TeV}$&0.5&2&\\
\hline
$\delta_{12}^{LL}$&0&0.05&\\
$\delta_{12}^{RR}$&0&0.05&\\
$\delta_{12}^{LR}$&0&0.002&\\
\hline
\end{tabular*}
\caption{Scanning parameters for Figs.~\ref{9}-\ref{11}.}\label{table2}
\end{table*}

\subsection{Scanning diagram of the effect of slepton mixing parameters  $\delta_{12}^{XX}~(X=L,R)$ on the $\mu-e$ conversion rate }

In the above subsections, we show only the effect of parameters on the $\mu-e$ conversion rate and ${\rm{Br}}(\mu\rightarrow e\gamma)$. In this subsection, we scan the parameter space shown in Table~\ref{table2}, in order to clearly see the constraints of ${\rm{Br}}(\mu\rightarrow e\gamma)$ on the $\mu-e$ conversion rate at more parameter space. Under the condition that the SM-like Higgs boson mass $m_h$\emph{=}$125.09\pm 0.24\: {\rm{GeV}}$ in 3$\sigma$, neutral fermion masses $m_{\chi_\eta^o}>200~{\rm GeV}$ $(\eta=1,\cdots,7)$ and the scalar masses $m_{S_{m, n}^{c}}>500~{\rm GeV}$ $(m,n=1,\cdots,6)$ are satisfied. By randomly scanning 20,000 points, we obtain the relation of the $\mu- e$ conversion rate in the different nuclei and ${\rm{Br}}(\mu\rightarrow e\gamma)$ versus $\delta_{12}^{XX}~(X=L,R)$, respectively. When the variable is $\delta_{12}^{XX}~(X=L,R)$ in the figures, the other two are $\delta_{12}^{XX}~(X=L,R)=0$.

\begin{figure}
\setlength{\unitlength}{1mm}
\centering
\includegraphics[width=2.8in]{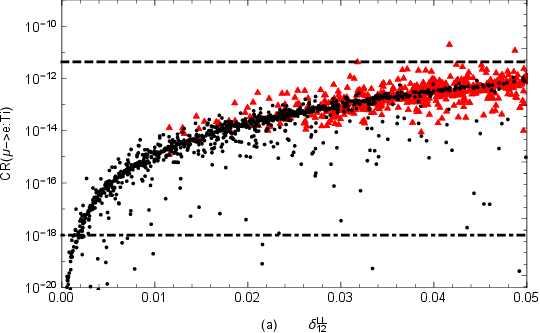}
\vspace{0cm}
\includegraphics[width=2.8in]{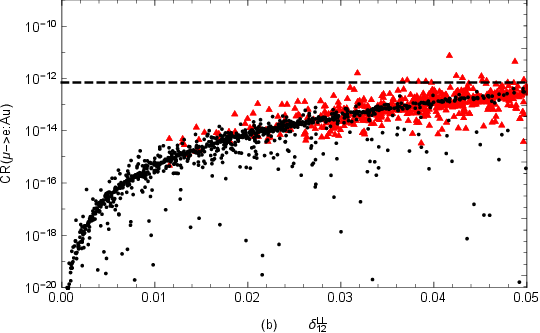}
\vspace{0cm}
\includegraphics[width=2.8in]{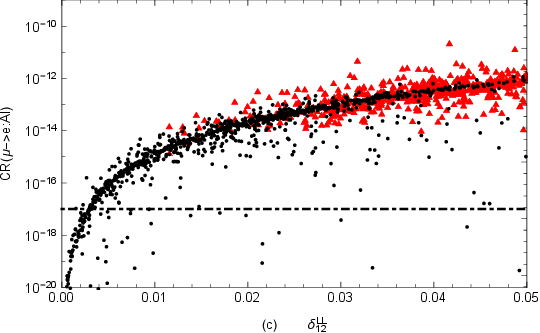}
\vspace{0cm}
\includegraphics[width=2.8in]{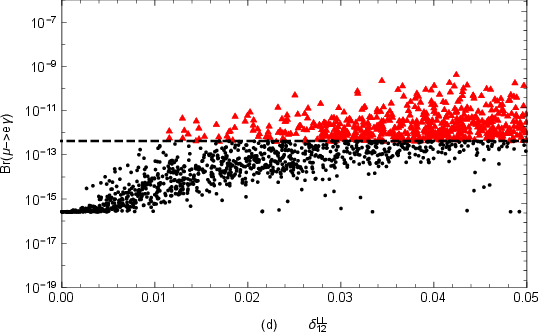}
\vspace{0cm}
\caption[]{{\label{9}} (a) ${\rm{CR}}(\mu\rightarrow e:\rm{Ti})$, (b) ${\rm{CR}}(\mu\rightarrow e:\rm{Au})$, (c) ${\rm{CR}}(\mu\rightarrow e:\rm{Al})$, and (d) ${\rm{Br}}(\mu \rightarrow e\gamma)$ versus $\delta_{12}^{LL}$ after randomly scanning Table~\ref{table2}, where the dashed and dot-dashed lines denote the present limits and future sensitivities respectively. Here, the red triangles are ruled out by the present limit of ${\rm{Br}}(\mu \rightarrow e\gamma)$, and the black dots are consistent with the present limit of ${\rm{Br}}(\mu \rightarrow e\gamma)$.}
\end{figure}

\begin{figure}
\setlength{\unitlength}{1mm}
\centering
\includegraphics[width=2.8in]{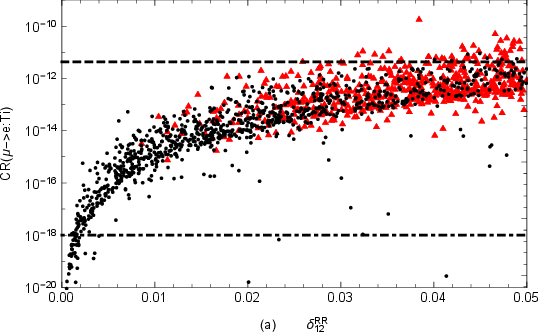}
\vspace{0cm}
\includegraphics[width=2.8in]{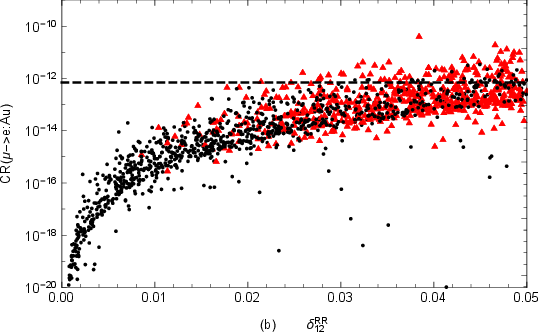}
\vspace{0cm}
\includegraphics[width=2.8in]{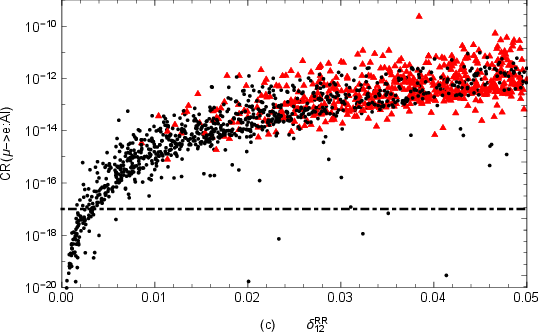}
\vspace{0cm}
\includegraphics[width=2.8in]{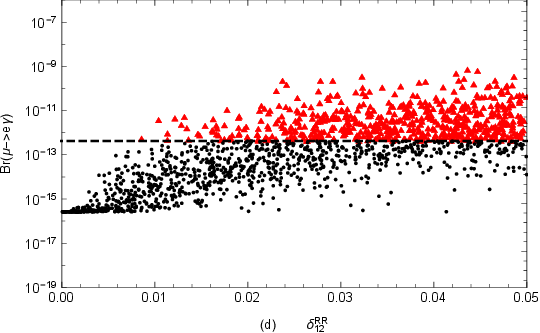}
\vspace{0cm}
\caption[]{{\label{10}} (a) ${\rm{CR}}(\mu\rightarrow e:\rm{Ti})$, (b) ${\rm{CR}}(\mu\rightarrow e:\rm{Au})$, (c) ${\rm{CR}}(\mu\rightarrow e:\rm{Al})$, and (d) ${\rm{Br}}(\mu \rightarrow e\gamma)$ versus $\delta_{12}^{RR}$ after randomly scanning Table~\ref{table2}.}
\end{figure}

\begin{figure}
\setlength{\unitlength}{1mm}
\centering
\includegraphics[width=2.8in]{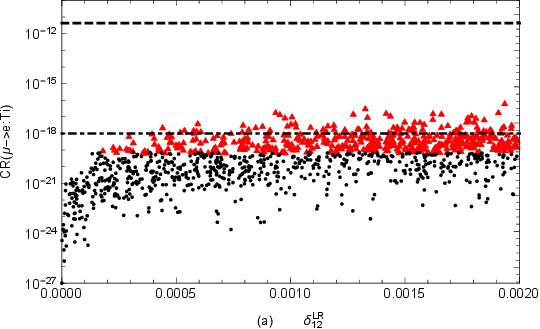}
\vspace{0cm}
\includegraphics[width=2.8in]{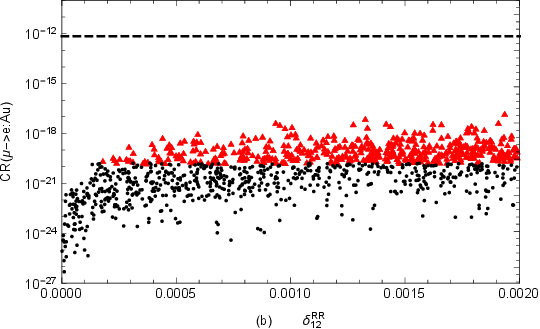}
\vspace{0cm}
\includegraphics[width=2.8in]{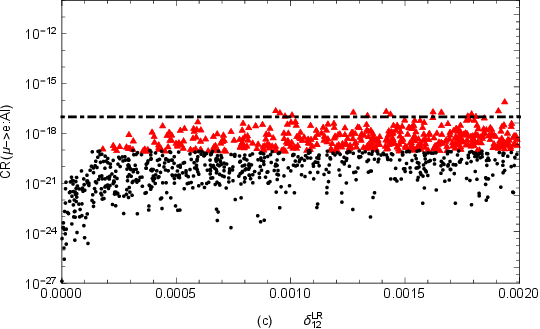}
\vspace{0cm}
\includegraphics[width=2.8in]{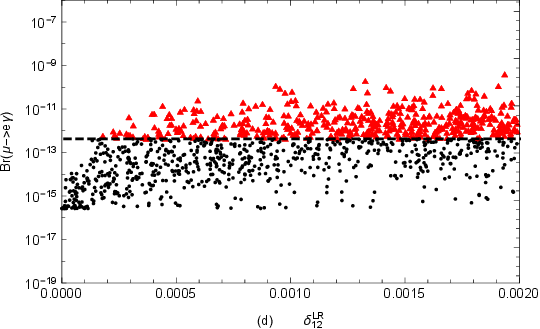}
\vspace{0cm}
\caption[]{{\label{11}} (a) ${\rm{CR}}(\mu\rightarrow e:\rm{Ti})$, (b) ${\rm{CR}}(\mu\rightarrow e:\rm{Au})$, (c) ${\rm{CR}}(\mu\rightarrow e:\rm{Al})$, and (d) ${\rm{Br}}(\mu \rightarrow e\gamma)$ versus $\delta_{12}^{LR}$ after randomly scanning Table~\ref{table2}.}
\end{figure}

In Fig.~\ref{9}, we plot ${\rm{CR}}(\mu\rightarrow e:\rm{Ti})$, ${\rm{CR}}(\mu\rightarrow e:\rm{Au})$, ${\rm{CR}}(\mu\rightarrow e:\rm{Al})$, and ${\rm{Br}}(\mu\rightarrow e\gamma)$ versus the slepton flavor mixing parameter $\delta_{12}^{LL}$, after randomly scanning Table~\ref{table2}.
By observing Fig.~\ref{9}, we can find that the constraint of ${\rm{Br}}(\mu \rightarrow e\gamma)$ to the $\mu - e$ conversion rate is relatively large. Looking at Fig.~\ref{9} (a), we find that ${\rm{CR}}(\mu\rightarrow e:\rm{Ti})$ can go to $\mathcal{O}(10^{-12})$ under the constraint of the upper limit of ${\rm{Br}}(\mu \rightarrow e\gamma)$. Although ${\rm{CR}}(\mu\rightarrow e:\rm{Ti})$ does not exceed the upper limit of the current experiment, it obviously exceeds the sensitivity of future experiments. In Fig.~\ref{9} (b), ${\rm{CR}}(\mu\rightarrow e:\rm{Au})$ can go beyond $\mathcal{O}(10^{-13})$ under the constraint of ${\rm{Br}}(\mu \rightarrow e\gamma)$. In Fig.~\ref{9} (c), ${\rm{CR}}(\mu\rightarrow e:\rm{Al})$ can exceed its sensitivity to future experiments under the constraint of ${\rm{Br}}(\mu \rightarrow e\gamma)$ and can also reach $\mathcal{O}(10^{-12})$. It is likely that the $\mu- e$ conversion rate in $_{22}^{48}{\rm{Ti}}$ and $_{13}^{27}{\rm{Al}}$ will be detected in the near future with increasing experimental accuracy.

We also plot ${\rm{CR}}(\mu\rightarrow e:\rm{Ti})$, ${\rm{CR}}(\mu\rightarrow e:\rm{Au})$, ${\rm{CR}}(\mu\rightarrow e:\rm{Al})$, and ${\rm{Br}}(\mu\rightarrow e\gamma)$ versus the slepton flavor mixing parameter $\delta_{12}^{RR}$  in Fig.~\ref{10}.
By observing the black spots that conform to the constraint of ${\rm{Br}}(\mu \rightarrow e\gamma)$ in Fig.~\ref{10}, we find that ${\rm{CR}}(\mu\rightarrow e:\rm{Ti})$ and ${\rm{CR}}(\mu\rightarrow e:\rm{Au})$ can achieve $\mathcal{O}(10^{-12})$ which can reach the upper limit of current experiments. In addition ${\rm{CR}}(\mu\rightarrow e:\rm{Al})$ can also attain $\mathcal{O}(10^{-12})$, which is 5 orders of magnitude larger than the future experimental sensitivity at the Mu2e and COMET experiments.

In Fig.~\ref{11}, the $\mu - e$ conversion rate and ${\rm{Br}}(\mu\rightarrow e\gamma)$ versus the slepton flavor mixing parameter $\delta_{12}^{LR}$ are plotted. The numerical results show that the upper limit of ${\rm{Br}}(\mu \rightarrow e\gamma)$ is very strict on the $\mu - e$ conversion rate in the different nuclei through the slepton flavor mixing parameter $\delta_{12}^{LR}$. By observing black dots in Fig.~\ref{11}, it can be found that under the constraint of ${\rm{Br}}(\mu \rightarrow e\gamma)$, the $\mu - e$ conversion rate in nuclei can be about $\mathcal{O}(10^{-19})$, which is under the sensitivity of future experiments. This means that the effect of the slepton mixing parameter $\delta_{12}^{LR}$ on the $\mu - e$ conversion rate in the different nuclei is small, constrained by the upper limit of ${\rm{Br}}(\mu \rightarrow e\gamma)$.

\section{Summary\label{sec5}}
In this work, we have studied the lepton flavor violating process of $\mu-e$ conversion in nuclei within the framework of the B-LSSM. The numerical results show that the $\mu-e$ conversion rate in nuclei depends on the slepton flavor mixing parameters $\delta_{12}^{XX}~(X=L,R)$ because the lepton flavor violating  processes are flavor dependent. Under the constraint of the experimental upper limit on the LFV branching ratio of $\mu\rightarrow e\gamma$, the $\mu-e$ conversion rate in $_{22}^{48}\rm{Ti}$ and  $_{\:79}^{197}{\rm{Au}}$ nuclei can attain $\mathcal{O}(10^{-12})$, which can reach the experimental upper limits. The $\mu-e$ conversion rate in $_{13}^{27}{\rm{Al}}$ nuclei can also reach $\mathcal{O}(10^{-12})$, which is 5 orders of magnitude larger than the future experimental sensitivity at the Mu2e and COMET experiments.

Compared with the MSSM, exotic two singlet Higgs fields and three generations of right-handed neutrinos in the B-LSSM induce new sources for the lepton flavor violation. Note that $\tan\beta'$ and $g_{YB}$ are new parameters in the B-LSSM beyond  the MSSM, which can affect the numerical results through the new mass matrix of sleptons, Higgs bosons, and neutralinos. Numerical results indicate that the new physics corrections dominate the evaluations on the $\mu-e$ conversion rates in nuclei in some parameter space of the B-LSSM. The theoretical predictions on the $\mu-e$ conversion rates in $_{22}^{48}\rm{Ti}$ and $_{13}^{27}{\rm{Al}}$ nuclei can easily exceed the future experimental sensitivities and may be detected in the near future.

\begin{acknowledgments}
\indent\indent
This work was supported by the National Natural Science Foundation of China with Grants No. 11705045, No. 12075074, and No. 11535002,  Natural Science Foundation for Distinguished Young Scholars of Hebei Province with Grant No. A2022201017, Natural Science Foundation of Guangxi Autonomous Region with Grant No. 2022GXNSFDA035068, Post-graduate's Innovation Fund Project of Hebei University under Grant No. HBU2022BS002, the Natural Science Foundation of Hebei Province under Grant No. A2020201002, the youth top-notch talent support program of the Hebei Province, and the Midwest Universities Comprehensive Strength Promotion project.
\end{acknowledgments}

\appendix

\section{loop function \label{form factors}}
The loop function $I_i$ and $G_i$ are written by

\begin{eqnarray}
&&{I_1}(\textit{x}_1 , x_2 ) = \frac{1}{{96{\pi ^2}}} \Big[ \frac{{11 + 6\ln {x_2}}}{{({x_2} - {x_1})}}- \frac{{15{x_2} + 18{x_2}\ln {x_2}}}{{{{({x_2} - {x_1})}^2}}} + \frac{{6x_2^2 + 18x_2^2\ln {x_2}}}{{{{({x_2} - {x_1})}^3}}}  \nonumber\\
&&\hspace{2.1cm} + \: \frac{{6x_1^3\ln {x_1}}-{6x_2^3\ln {x_2}}}{{{{({x_2} - {x_1})}^4}}}  \Big], \\
&&{I_2}(\textit{x}_1 , x_2 ) = \frac{1}{{16{\pi ^2}}}\Big[ \frac{{1 + \ln {x_2}}}{{({x_2} - {x_1})}} + \frac{{{x_1}\ln {x_1}}-{{x_2}\ln {x_2}}}{{{{({x_2} - {x_1})}^2}}} \Big],\\
&&{I_3}(\textit{x}_1 , x_2 ) = \frac{1}{{32{\pi ^2}}}\Big[  \frac{{3 + 2\ln {x_2}}}{{({x_2} - {x_1})}} - \frac{{2{x_2} + 4{x_2}\ln {x_2}}}{{{{({x_2} - {x_1})}^2}}} -\frac{{2x_1^2\ln {x_1}}}{{{{({x_2} - {x_1})}^3}}}  +  \frac{{2x_2^2\ln {x_2}}}{{{{({x_2} - {x_1})}^3}}}\Big], \\
&& {G_1}(\textit{x}_1 , x_2 , x_3) \nonumber\\
&&\hspace{0.6cm} =  \frac{1}{{16{\pi ^2}}}\Big[  \frac{{x_1^2\ln {x_1}}}{{({x_1} - {x_2})({x_1} - {x_3})}} +\frac{{x_2^2\ln {x_2}}}{{({x_2} - {x_1})({x_2} - {x_3})}} +   \frac{{x_3^2\ln {x_3}}}{{({x_3} - {x_1})({x_3} - {x_2})}} \Big], \quad \\
&& {G_2}(\textit{x}_1 , x_2 , x_3, x_4) \nonumber\\
&&\hspace{0.6cm} = \frac{1}{{16{\pi ^2}}}\Big[\frac{{{x_1}\ln {x_1}}}{{({x_1} - {x_2})({x_1} - {x_3})({x_1} - {x_4})}}  + \frac{{{x_2}\ln {x_2}}}{{({x_2} - {x_1})({x_2} - {x_3})({x_2} - {x_4})}} \nonumber\\
&&\hspace{0.6cm}\quad + \frac{{{x_3}\ln {x_3}}}{{({x_3}  - {x_1})({x_3} - {x_2})({x_3} - {x_4})}} + \: \frac{{{x_4}\ln {x_4}}}{{({x_4} - {x_1})({x_4} - {x_2})({x_4} - {x_3})}}\Big] , \\
&&{G_3}(\textit{x}_1 , x_2 , x_3, x_4) \nonumber\\
&&\hspace{0.6cm} = \frac{1}{{16{\pi ^2}}}\Big[\frac{{x_1^2\ln {x_1}}}{{({x_1} - {x_2})({x_1} - {x_3})({x_1} - {x_4})}}   + \frac{{x_2^2\ln {x_2}}}{{({x_2} - {x_1})({x_2} - {x_3})({x_2} - {x_4})}} \nonumber\\
&&\hspace{0.6cm}\quad + \frac{{x_3^2\ln {x_3}}}{{({x_3}  - {x_1})({x_3} - {x_2})({x_3} - {x_4})}}   +  \frac{{x_4^2\ln {x_4}}}{{({x_4} - {x_1})({x_4} - {x_2})({x_4} - {x_3})}}\Big].
\end{eqnarray}

\section{Couplings \label{app-coupling}}
The coupling $C$ can be written as

\begin{eqnarray}
&&C_L^{S_m^c \chi _\eta ^o{{\bar l }_{i}}}=C_L^{S_m^c \chi _\sigma ^o{{\bar l }_{i}}}=-\sqrt2g_1N_{\eta1}^*\sum_{a=1}^3Z_{m(3+a)}^{E,*}U_{R,ia}^{e,*}\nonumber\\
&&\hspace{3.5cm}-1/\sqrt2(2g_{YB}+g_B)N_{\eta5}^*\sum_{a=1}^3Z_{m(3+a)}^{E,*}U_{R,ia}^{e,*}\nonumber\\
&&\hspace{3.5cm}-N_{\eta3}^*\sum_{b=1}^3Z_{mb}^{E,*}\sum_{a=1}^3U_{R,ia}^{e,*}Y_{e,ab},\\
&&C_R^{S_m^c \chi _\eta ^o{{\bar l }_{i}}}=1/2(-2\sum_{b=1}^3\sum_{a=1}^3Y_{e,ab}^*Z_{m(3+a)}^{E,*}U_{L,ib}^eN_{\eta3}+\sqrt2\sum_{a=1}^3Z_{ma}^{E,*}U_{L,ia}^e(g_1N_{\eta1}+g_2N_{\eta2}\nonumber\\
&&\hspace{1.5cm}+(g_{YB}+g_B)N_{\eta5})),\\
&&C_L^{S_m^{c*} {l_{j}}\bar \chi _\eta ^o}=1/2(\sqrt2g_1N_{\eta1}^*\sum_{a=1}^3U_{L,ja}^{e,*}Z_{ma}^E+\sqrt2g_2N_{\eta2}^*\sum_{a=1}^3U_{L,ja}^{e,*}Z_{ma}^E
+\sqrt2g_{YB}N_{\eta5}^*\sum_{a=1}^3U_{L,ja}^{e,*}Z_{ma}^E\nonumber\\
&&\hspace{1.7cm}+\sqrt2g_BY_{\eta5}^*\sum_{a=1}^3U_{L,ja}^{e,*}Z_{ma}^E-2N_{\eta3}^*\sum_{b=1}^3U_{L,jb}^{e,*}\sum_{a=1}^3Y_{e,ab}Z_{m(3+a)}^E),\\
&&C_R^{S_m^{c*} {l_{j}}\bar \chi _\eta ^o}=-1/\sqrt2\sum_{a=1}^3Z_{m(3+a)}^EU_{R,ja}^e(2g_1N_{\eta1}+(2g_{YB}+g_B)N_{\eta5})-\sum_{b=1}^3\sum_{a=1}^3Y_{e,ab}^*U_{R,ja}^eZ_{mb}^EN_{\eta3},\nonumber\\
&&\hspace{1.7cm} \;\;\;\quad\\
&&C_L^{Z S_m^{c} S_n^{c*}}=1/2(2g_1 cos\Theta_w \sum_{a=1}^3 Z_{m3+a}^{E,*}Z_{n3+a}^{E}+(g_1 cos\Theta_W+g_2sin\Theta_W)\sum_{a=1}^3Z_{ma}^{E,*}Z_{na}^{E}),\\
&&C_L^{{\tilde{q}_I} \chi _\sigma^0{{\bar q }_i}\ast}=C_L^{{\tilde{q}_I} \chi _\eta^0{{\bar q }_i}\ast}=-1/6 (\sqrt2 g_1 N_{\sigma1}^*\sum_{a=1}^3U_{L,ia}^{q,*}Z_{Ia}^Q-3\sqrt2 g_2N_{\sigma2}^*\sum_{a=1}^3U_{L,ia}^{q,*}Z_{Ia}^Q\nonumber\\
&&\hspace{3.5cm}+\sqrt2 g_{YB}N_{\sigma5}^*\sum_{a=1}^3U_{L,ia}^{q,*}Z_{Ia}^Q+\sqrt2 g_B N_{\sigma5}^*\sum_{a=1}^3U_{L,ia}^{q,*}Z_{Ia}^Q\nonumber\\
&&\hspace{3.5cm}+6 N_{\sigma3}^*\sum_{b=1}^3U_{L,ib}^{q,*}\sum_{a=1}^3Y_{q,ab}Z_{I3+a}^Q)\\
&&C_R^{{\tilde{q}_I} \chi _\sigma^0{{\bar q }_i}\ast}=C_R^{{\tilde{q}_I} \chi _\eta^0{{\bar q }_i}\ast}=1/6((-6\sum_{b=1}^3\sum_{a=1}^3Y_{q,ab}^*U_{R,ia}^qZ_{Ib}^QN_{\sigma3}\nonumber\\
&&\hspace{3.5cm}+\sqrt2\sum_{a=1}^3Z_{I3+a}^QU_{R,ia}^q(-2g_1N_{\sigma1}+(-2g_{YB}+g_B)N_{\sigma5}))\\
&&C_L^{{\tilde{q}_I} \chi _\sigma^0{{\bar q }_i}}=C_L^{{\tilde{q}_I} \chi _\eta^0{{\bar q }_i}}=1/6(\sqrt2(-2 g_1)N_{\sigma1}^*\sum_{a=1}^3Z_{I3+a}^{Q,*}U_{R,ia}^{q,*}\nonumber\\
&&\hspace{3.5cm}+\sqrt2(-2 g_{YB}+g_B)N_{\sigma5}^*\sum_{a=1}^3Z_{I3+a}^{Q,*}U_{R,ia}^{q,*}\nonumber\\
&&\hspace{3.5cm}-6N_{\sigma3}^*\sum_{b=1}^3Z_{Ib}^{Q,*}\sum_{a=1}^3U_{R,ia}^{q,*}Y_{q,ab})\\
&&C_R^{{\tilde{q}_I} \chi _\sigma^0{{\bar q }_i}}=C_R^{{\tilde{q}_I} \chi _\eta^0{{\bar q }_i}}=-1/6(6 \sum_{b=1}^3\sum_{a=1}^3Y_{q,ab}^*Z_{I3+a}^{Q,*}U_{L,ib}^qN_{\sigma3}\nonumber\\
&&\hspace{3.5cm}+\sqrt2\sum_{a=1}^3Z_{Ia}^{Q,*}U_{L,ia}^q(-3g_2N_{\sigma2}+g_1N_{\sigma1}+(g_{YB}+g_B))N_{\sigma5})).
\end{eqnarray}

\end{document}